# MATERIALS ASPECTS OF HIGH-TEMPERATURE SUPERCONDUCTORS FOR APPLICATIONS

**Roland HOTT**
Forschungszentrum Karlsruhe, Institut für Festkörperphysik,
P. O. Box 3640, 76021 Karlsruhe, GERMANY
Email: roland.hott@ifp.fzk.de

## 1. Introduction

Sixteen years after their discovery [1], cuprate high-temperature superconductors (HTS) are still one of the topical centers of interest in physics. The mechanism of superconductivity is still under lively debate [2,3,4]. The description of their normal state properties [5,6] has turned out to represent an even bigger challenge to Solid State Physics theory. A major part of the progress in clarifying these questions is due to the improvement of sample quality [7,8,9]. This may be attributed in part to the efforts of Applied Superconductivity to create a materials basis for the multitude of applications, which were envisioned in the unprecedented scientific euphoria raised by the HTS discovery[1].

The present situation of HTS materials science resembles in many aspects the history of semiconductor materials development half a century ago [11]. However, in sharp contrast in particular to the case of silicon, HTS are *multi-element* compounds based on complicated sequences of *oxide layers* [12,13,14]. In addition to the impurity problem due to undesired additional elements which gave early semiconductor research a hard time in establishing reproducible materials properties, intrinsic local stochiometry defects arise in HTS from the insertion of cations in the wrong layer and defects of the oxygen sublattice. As additional requirement for the optimization of the superconducting properties, the oxygen content has to be adjusted in a compound-specific off-stochiometric ratio [15], but nevertheless with a spatially homogeneous microscopic distribution of the resulting oxygen vacancies or interstitials [16].

[1] It has been estimated that in the hottest days of this "Woodstock of physics" in 1987 [10] about one third of *all* physicists all over the world tried to contribute to the HTS topic. Meanwhile, the number of publications on HTS has increased to > 100 000.

The discovery of High-Temperature Superconductivity has thus challenged research in a class of complicated compounds which otherwise would have been encountered on the classical research route of systematic investigation with gradual increase of materials complexity only in a far future. In the quest for better HTS materials quality, the very short superconducting coherence length of the order of the dimensions of the crystallographic unit cell did not allow easy progress. On the other hand, this experimental restriction leads to a close connection of macroscopically measurable superconducting properties with micro-structural details which permits a kind of atomic scale monitoring of the HTS materials quality. The huge efforts in HTS materials science have thus paved the way for the improved preparation of oxide materials in general, e. g. ferroelectric oxides [17] or manganites [18,19]. The broad spectrum of unusual physical properties of this new generation of "sophisticated oxides" has opened new perspectives for fundamental as well as for applied research [20,21].

Today, reproducible preparation techniques for a number of HTS material species are available which provide a first materials basis for applications. As a stroke of good fortune, the optimization of these materials with respect to their superconducting properties seems to be in accord with the efforts to improve their stability in technical environments in spite of the only metastable chemical nature of these substances under such conditions [22,23,24].

## 2. Atomic Structure and Classification

The structural element of HTS compounds related to the location of mobile charge carriers are stacks of a certain number n = 1, 2, 3, ... of $CuO_2$ layers which are "glued" on top of each other by means of intermediate Ca layers (See Fig.1&2) [12,13,25,26]. Counterpart of these "*active blocks*" of $(CuO_2/Ca/)_{n-1}CuO_2$ stacks are "*charge reservoir blocks*" $EO/(AO_x)_m/EO$ with m = 1, 2 monolayers of a quite arbitrary oxide $AO_x$ (A = Bi [13], Pb [25], Tl [13], Hg [13], Au [27], Cu [13], Ca [28], B [25], Al [25], Ga [25]; see Table 1)[2] "wrapped" on each side by a monolayer of alkaline earth oxide EO with E = Ba, Sr (see Fig. 1&2). The HTS structure results from alternating stacking of these two block units. The choice of BaO or SrO as "wrapping" layer is not arbitrary but depends on the involved $AO_x$ since it has to provide a good spatial adjustment of the $CuO_2$ to the $AO_x$ layers.

The general chemical formula[3] $\mathbf{A_m}E_\mathbf{2}Ca_{\mathbf{n-1}}Cu_\mathbf{n}O_{2n+m+2+y}$ (see Fig. 2) is conveniently abbreviated as **A-m2(n-1)n** [26] (e. g. $Bi_2Sr_2Ca_2Cu_3O_{10}$: Bi-2223) neglecting the indication of the alkaline earth element[4] (see Tab.1). The family of all n = 1, 2, 3,... representatives with common $AO_x$ will be referred here as “A-HTS”, e. g. Bi-HTS.

The most prominent compound $\mathbf{Y}Ba_2\mathbf{Cu_3O_7}$ (see Fig. 1), the first HTS discovered with a critical temperature $T_c$ for the onset of superconductivity above the boiling point of liquid nitrogen [29], is traditionally abbreviated as "**YBCO**" or "**Y-123**" ($\mathbf{Y_1}Ba_\mathbf{2}Cu_\mathbf{3}O_{7-\delta}$). It also fits into the general HTS classification scheme as a modification of Cu-1212 where Ca is completely substituted by Y. This substitution introduces extra charge next to the $CuO_2$ layers due to the higher valence of Y (+3) compared to Ca (+2). The HTS compounds $RBa_2Cu_3O_{7-\delta}$ ("RBCO", "R-123") where R can be La [13] or any rare earth element [30] except for Ce or Tb [31] can be regarded as a generalization of this substitution scheme. The lanthanide contraction of the R ions provides an experimental handle on the distance between the two $CuO_2$ layers of the active block of the doped Cu-1212 compound [32]. $\mathbf{Y_1Ba_2Cu_4O_8}$ ("**Y-124**") is the m = 2 counterpart Cu-2212 of YBCO. The "**214**" HTS compounds $E_2Cu_1O_4$ (s. Tab1), e. g. $La_{2-x}Sr_xCuO_4$ ("LSCO") or $Nd_{2-x}Ce_xCuO_4$ ("NCCO") are a bit exotic in this ordering scheme but may also be represented here as “0210” with m = 0, n = 1 and $E_2 = La_{2-x}Sr_x$ and $E_2 = Nd_{2-x}Ce_x$, respectively[5].

[2] HTS compounds based on “oxide mixtures” $A^{(1)}_s A^{(2)}_{1-s} O_x$ are neglected in this consideration.
[3] y = m (x – 1) in the oxygen stoichiometry factor.
[4] A more precise terminology $A\text{-}m^{(E)}2(n\text{-}1)n$ has been suggested [26].

Further interesting chemical modifications of the basic HTS compositions are the introduction of fluorine [33,34] or chlorine [35] as more electronegative substituents of oxygen. This often results in a slight improvement of $T_c$: For Hg-1223, $T_c$ = 135 K can thus be raised to 138 K [36], the highest $T_c$ reported by now under normal pressure conditions (164 K at 30 GPa [37]). Another interesting modification which is realized for many HTS compounds are oxycarbonate HTS [38,39,40]. The inclusion of carbon into the HTS structure[6] with neutral or even slightly beneficial effect on the superconducting properties [40] is amazing since carbon impurities as secondary phases in HTS samples are usually very detrimental in this respect.

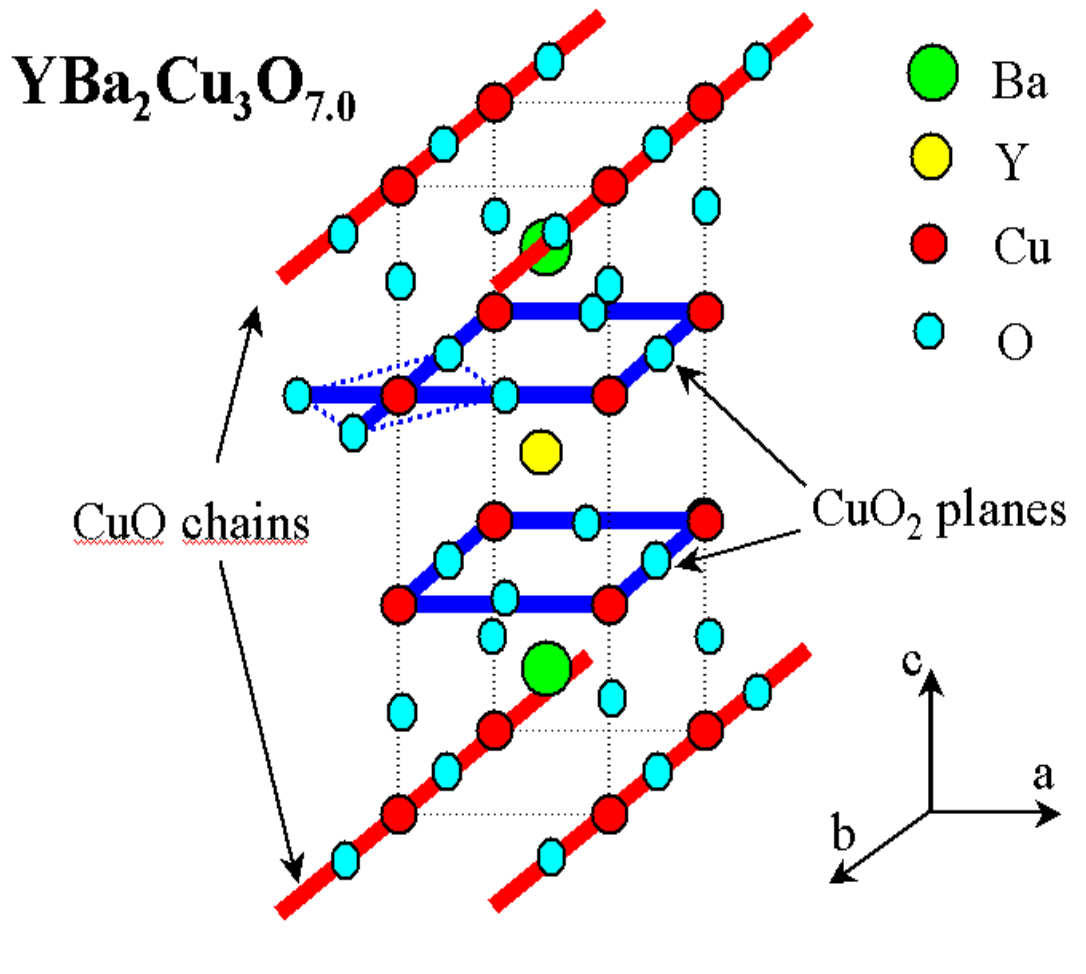

**Fig. 1** Crystal structure of $YBa_2Cu_3O_7$ ("YBCO"). The presence of the CuO chains introduces an orthorhombic distortion of the unit cell (a = 0.382 nm, b = 0.389 nm, c = 1.167 nm [65]).

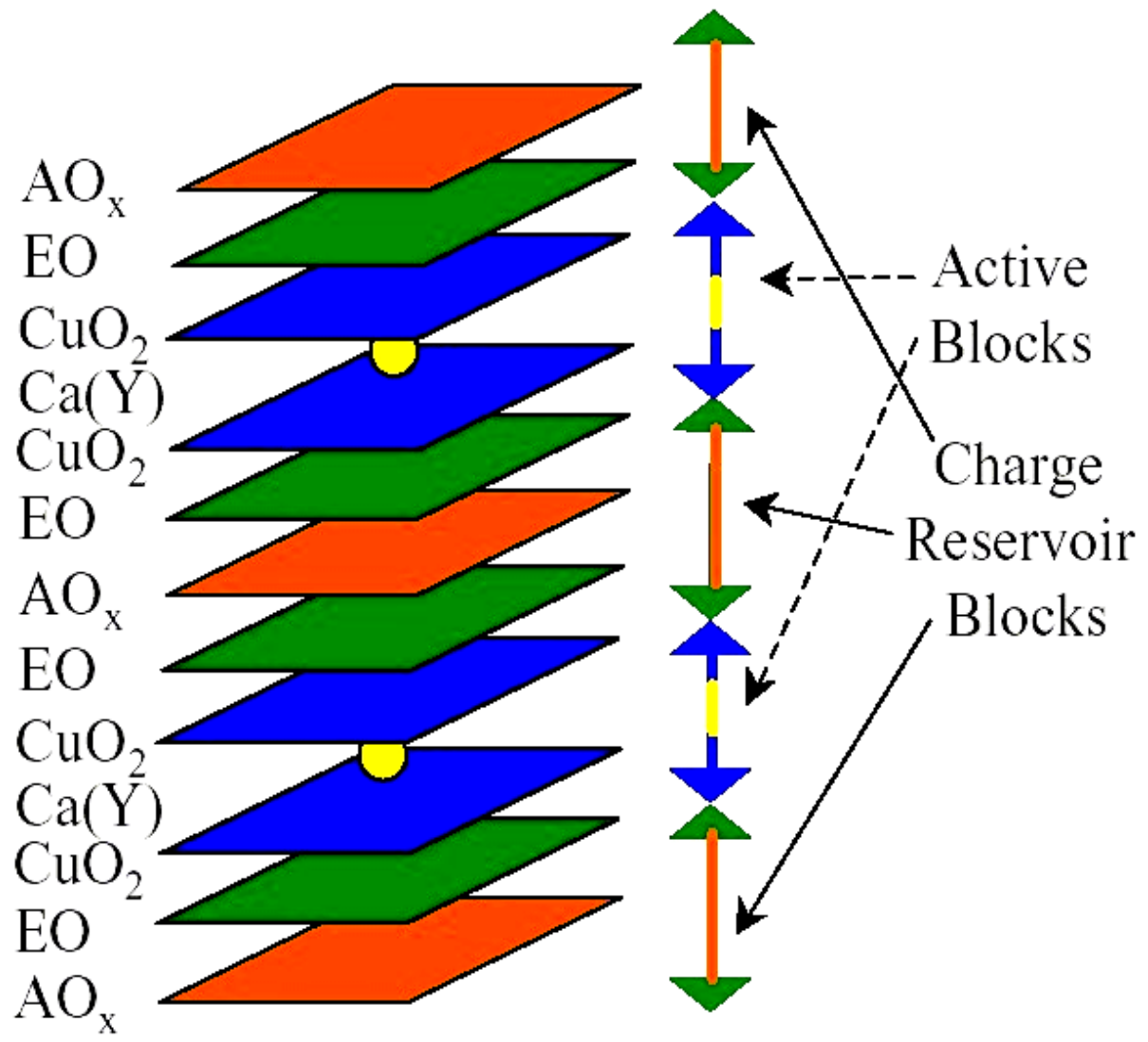

**Fig. 2** General structure of a cuprate HTS A-m2(n-1)n ($A_mE_2Ca_{n-1}Cu_nO_{2n+m+2+y}$) for m = 1. For m = 0 (or 2) the missing (additional) $AO_x$ layer per unit cell leads to a (a/2, b/2, 0)"side step" of the unit cells adjoining in c-axis direction.

[5] The missing BaO / SrO wrapping layer apparently leads to a much more sensitive behaviour of the superconducting properties with respect to structural disorder next to the $CuO_2$ layers.

[6] The carbon atoms are integrated as $CO_3$ units into the lattice structure of the $AO_x$ layer.

| HTS Family | Stochiometry | Notation | Compounds | Highest $T_c$ |
|---|---|---|---|---|
| Bi-HTS | $Bi_mSr_2Ca_{n-1}Cu_nO_{2n+m+2}$ m = 1, 2 n = 1, 2, 3 . . . | Bi-m2(n-1)n, BSCCO | Bi-1212 | 102 K [41] |
| | | | Bi-2201 | 34 K [42] |
| | | | Bi-2212 | 90 K [43] |
| | | | Bi-2223 | 110 K [13] |
| | | | Bi-2234 | 110 K [44] |
| Pb-HTS | $Pb_mSr_2Ca_{n-1}Cu_nO_{2n+m+2}$ | Pb-m2(n-1)n | Pb-1212 | 70 K [45] |
| | | | Pb-1223 | 122 K [46] |
| Tl-HTS | $Tl_mBa_2Ca_{n-1}Cu_nO_{2n+m+2}$ m = 1, 2 n = 1, 2, 3 . . . | Tl-m2(n-1)n, TBCCO | Tl-1201 | 50 K [13] |
| | | | Tl-1212 | 82 K [13] |
| | | | Tl-1223 | 133 K [47] |
| | | | Tl-1234 | 127 K [48] |
| | | | Tl-2201 | 90 K [13] |
| | | | Tl-2212 | 110 K [13] |
| | | | Tl-2223 | 128 K [49] |
| | | | Tl-2234 | 119 K [50] |
| Hg-HTS | $Hg_mBa_2Ca_{n-1}Cu_nO_{2n+m+2}$ m = 1, 2 n = 1, 2, 3 . . . | Hg-m2(n-1)n, HBCCO | Hg-1201 | 97 K [13] |
| | | | Hg-1212 | 128 K [13] |
| | | | Hg-1223 | 135 K [51] |
| | | | Hg-1234 | 127 K [51] |
| | | | Hg-1245 | 110 K [51] |
| | | | Hg-1256 | 107 K [51] |
| | | | Hg-2212 | 44 K [52] |
| | | | Hg-2223 | 45 K [53] |
| | | | Hg-2234 | 114 K [53] |
| Au-HTS | $Au_mBa_2Ca_{n-1}Cu_nO_{2n+m+2}$ | Au-m2(n-1)n | Au-1212 | 82 K [27] |
| 123-HTS | $RBa_2Cu_3O_{7-\delta}$ R = Y, La, Pr, Nd, Sm, Eu, Gd, Tb, Dy, Ho, Er, Tm, Yb, Lu | R-123, RBCO | Y-123, YBCO | 92 K [32] |
| | | | Nd-123, NBCO | 96 K [32] |
| | | | Gd-123 | 94 K [54] |
| | | | Er-123 | 92 K [7] |
| | | | Yb-123 | 89 K [30] |
| Cu-HTS | $Cu_mBa_2Ca_{n-1}Cu_nO_{2n+m+2}$ m = 1, 2 n = 1, 2, 3 . . . | Cu-m2(n-1)n | Cu-1223 | 60 K [13] |
| | | | Cu-1234 | 117 K [55] |
| | | | Cu-2223 | 67 K [13] |
| | | | Cu-2234 | 113 K [13] |
| | | | Cu-2245 | < 110 K [13] |
| Ru-HTS | $RuSr_2GdCu_2O_8$ | Ru-1212 | Ru-1212 | 72 K [56] |
| B-HTS | $B_mSr_2Ca_{n-1}Cu_nO_{2n+m+2}$ | B-m2(n-1)n | B-1223 | 75 K [57] |
| | | | B-1234 | 110 K [57] |
| | | | B-1245 | 85 K [57] |
| 214-HTS | $E_2CuO_4$ | LSCO | $La_{2-x}Sr_xCuO_4$ | 51 K [58] |
| | | "0201" | $Sr_2CuO_4$ | 25 (75)K [59] |
| | | *Electron-Doped HTS* | $La_{2-x}Ce_xCuO_4$ | 28 K [60] |
| | | PCCO | $Pr_{2-x}Ce_xCuO_4$ | 24 K [61] |
| | | NCCO | $Nd_{2-x}Ce_xCuO_4$ | 24 K [61] |
| | | | $Sm_{2-x}Ce_xCuO_4$ | 22 K [62] |
| | | | $Eu_{2-x}Ce_xCuO_4$ | 23 K [62] |
| | $Ba_2Ca_{n-1}Cu_nO_{2n+2}$ | "02(n-1)n" | "0212" | 90K [63] |
| | | | "0223" | 120K [63] |
| | | | "0234" | 105K [63] |
| | | | "0245" | 90K [63] |
| *Infinite-Layer* HTS | $ECuO_2$ | *Electron-Doped I.L.* | $Sr_{1-x}La_xCuO_2$ | 43 K [64] |

**Table 1** Classification and reported $T_c$ values of HTS compounds.

## 3. Structural Obstacles for Supercurrents

The High-$T_c$ euphoria in 1987 arrived soon at a first hangover after the first measurements of critical current densities: The $J_c$ values of only several 100 A/cm$^2$ obtained in the best polycrystalline material available at that time were below the current density level of copper wires at room-temperature at negligible energy dissipation. The situation was even worse since even weak magnetic fields turned out to be sufficient to kill these low supercurrents. This did not look like a continuation of the success story of LTS materials where magnet applications represent nowadays by far the biggest market and provide the basis of a substantial wire industry. Meanwhile, this "weak-link behavior" of HTS has been analyzed in detail and can now be overcome by special processing techniques.

Electrical currents create magnetic fields which tend to suppress superconductivity. This prevented for a long time high-current applications of superconductors. A first step towards this goal was the discovery of type-II superconductors where the magnetic penetration depth $\lambda$ is substantially longer than the superconducting coherence length $\xi$ and allows therefore the penetration of magnetic fields into the superconducting bulk. This coexistence of magnetic fields and superconductivity leads to a substantial reduction of the loss of superconductive condensation energy that has to be paid for magnetic field penetration and enables the survival of superconductivity even in strong magnetic fields. Fortunately, HTS are extreme type-II superconductors with $\lambda > 100$ nm and $\xi \sim 1$ nm.

Superconductivity in HTS is believed to have its origin in the physics of the $CuO_2$ layers where the mobile charges are located[7] [2,3]. The superconductive coupling between these $CuO_2$ layers within a given $(CuO_2/Ca/)_{n-1}CuO_2$ stack ("*interlayer coupling*") is much weaker than the *intralayer coupling* within the $CuO_2$ layers, but still much stronger than the coupling between the $(CuO_2/Ca/)_{n-1}CuO_2$ stacks which can be described as Josephson coupling (see Fig. 3). This quasi-2-dimensional nature of superconductivity in HTS leads to a pronounced anisotropy of the superconducting properties with much higher supercurrents along the $CuO_2$ planes than in the perpendicular direction, a property which is not at all appreciated with respect to technical applications but can be compensated by additional engineering efforts.

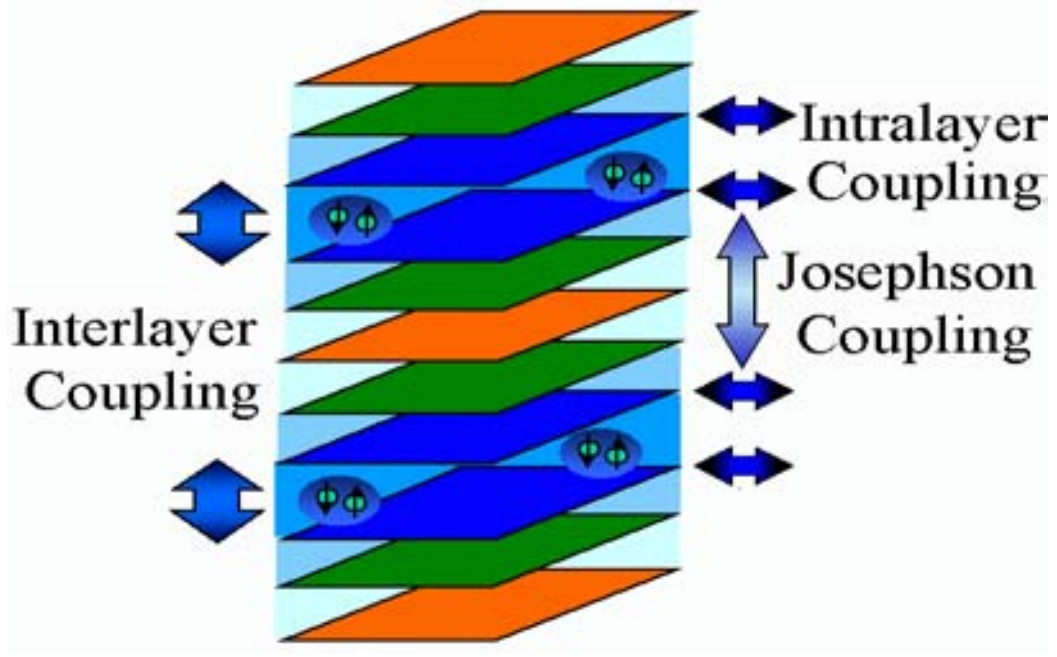

**Fig. 3** Hierarchy of the superconducting coupling between the different structural elements of cuprate HTS.

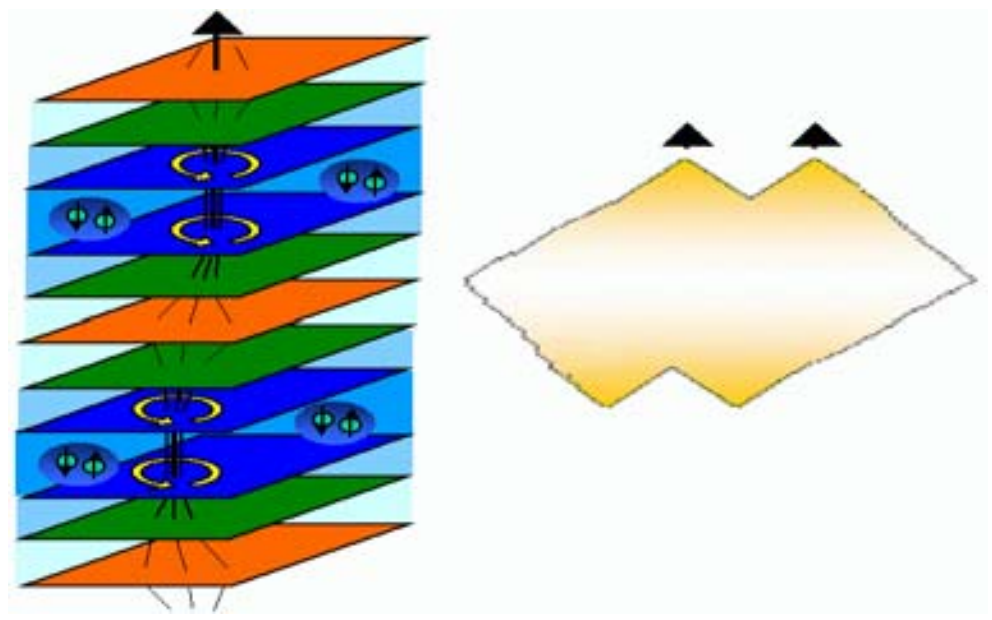

**Fig. 4** Quasi-disintegration of magnetic vortex lines into pancake vortices due to weak superconducting interlayer coupling and schematic overlap of neighboring vortices.

---

[7] In the present theoretical picture, the charge reservoir blocks play only a passive role in providing the doping charge as well as the storage space for extra oxygen ions and cations introduced on doping.

However, being type-II superconductor is not enough for a flow of strong currents without dissipation: The magnetic vortices that are introduced into the superconducting material by the magnetic field need to be fixed by *pinning centers,* or else dissipation by the *flux flow* of the vortices is induced. In "technical" superconductors, material imperfections of the dimension of the coherence length do this job by blocking superconductivity from these regions which provide then a natural "parking area" for vortex cores.

Material imperfections of the dimension of the coherence length are easily encountered in HTS due to their small coherence lengths, e. g. for YBCO values of $\xi_{ab} = 1.6$ nm, $\xi_c = 0.3$ nm for $T \rightarrow 0$ K [66] which are already comparable to the lattice parameters (YBCO: $a = 0.382$ nm, $b = 0.389$ nm, $c = 1.167$ nm [65]). However, the low $\xi_c$, i.e. the weak superconductive coupling between the $(CuO_2/Ca/)_{n-1}CuO_2$ stacks causes new problems. The thickness of the charge reservoir blocks $EO/(AO_x)_m/EO$ in-between these stacks is larger than $\xi_c$ with the result that due to the low Cooper pair density vortices are here no longer well-defined (see Fig. 4).

This leads to a quasi-disintegration of the vortices into stacks of "*pancake vortices*" which are much more flexible entities than the continuous quasi-rigid vortex lines in conventional superconductors and therefore require individual pinning centers. The extent of this quasi-disintegration is different for the various HTS compounds since $\xi_c$ is on the order of the thickness of a single oxide layers: Hence the number of layers in the charge reservoir blocks $EO/(AO_x)_m/EO$ makes a significant difference with respect to the pinning properties and thus to their supercurrents in magnetic fields. This is one of the reasons why YBCO ("Cu-**1**212") has a higher supercurrent capability in magnetic fields than the Bi-HTS Bi-**2**212 and Bi-**2**223 which for manufacturing reasons are still the most prominent HTS conductor materials[8].

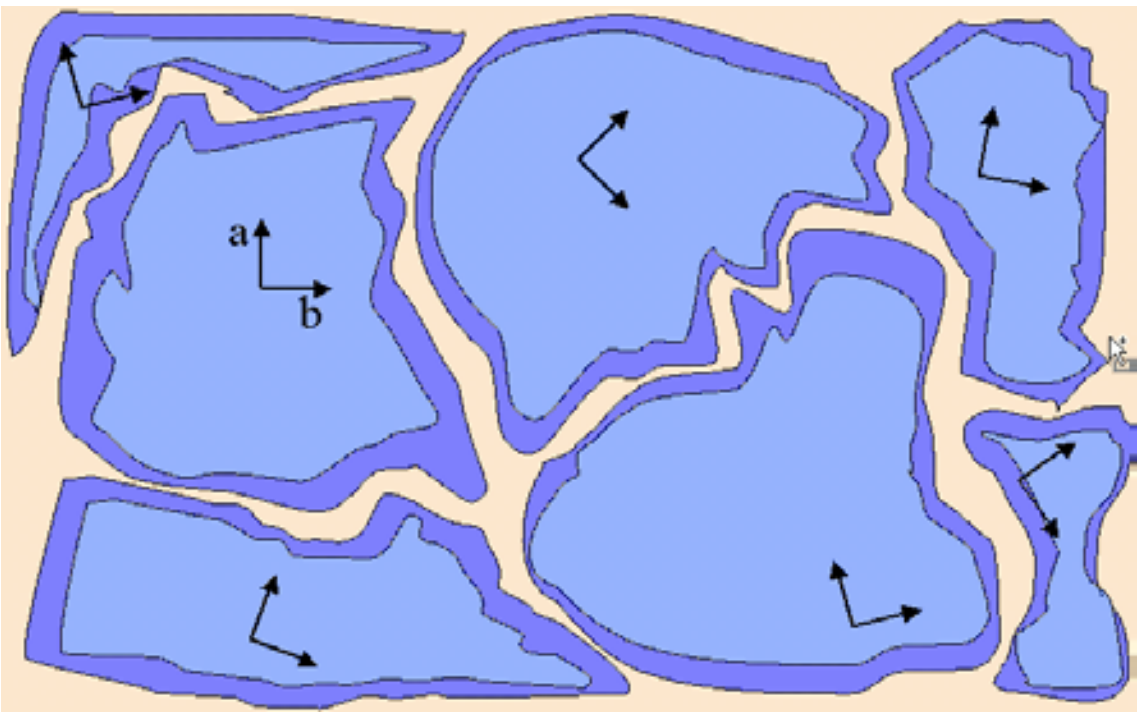

**Fig. 5** Schematics of the HTS microstructure: Differently oriented single crystal grains are separated by regions filled with secondary phases. In addition, oxygen depletion may occur at grain boundaries.

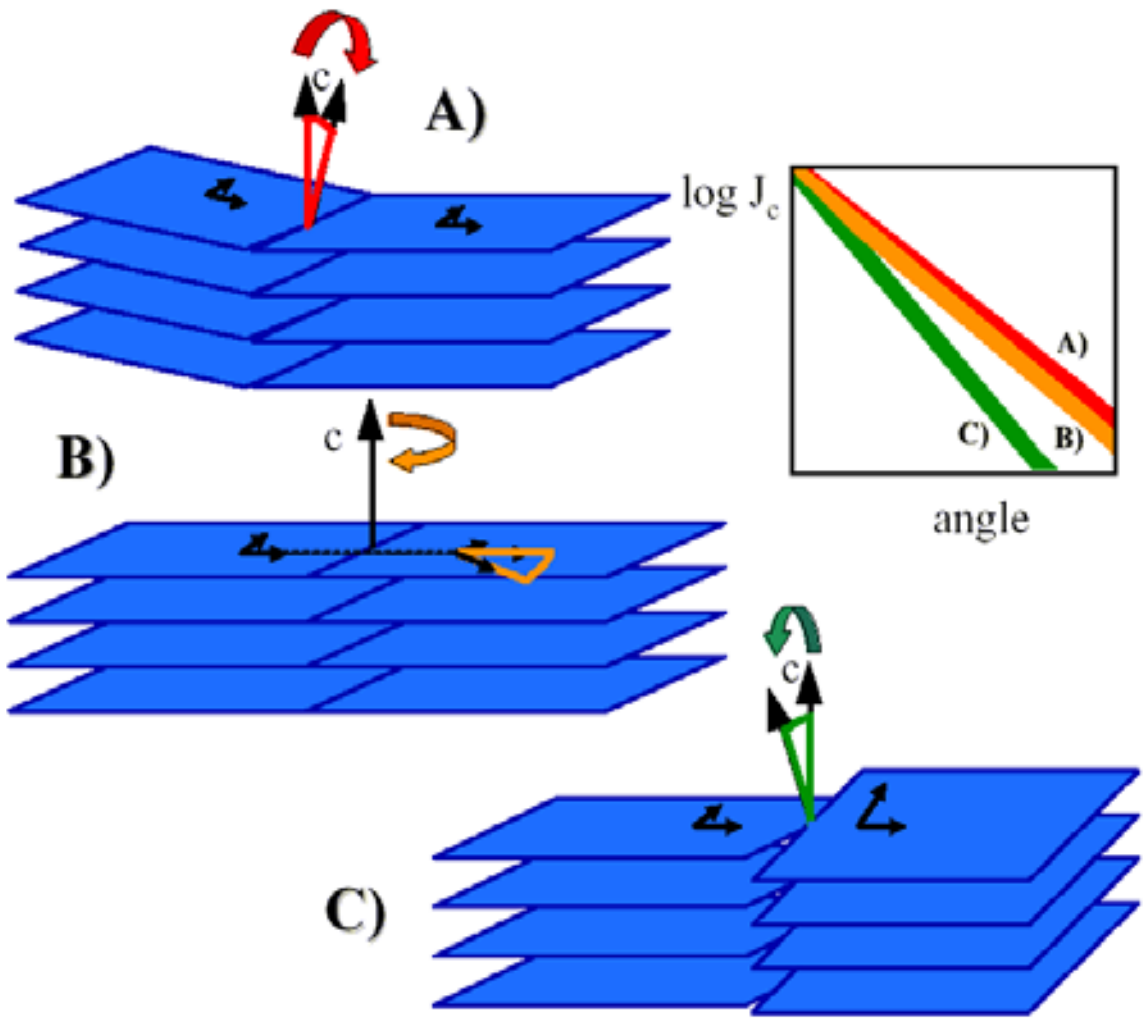

**Fig. 6** Basic grain boundary geometries and experimentally observed $J_c$ reduction $J_c \sim e^{-\alpha/\alpha_0}$ as function of the misalignment angle $\alpha$: $\alpha_0 \approx 5°$ for A) and B), $\alpha_0 \approx 3°$ for C) independent of temperature [71,72].

[8] In addition, in the Cu-HTS family the $AO_x$ layer is formed by CuO chain structures (see Fig. 1). There are indications that the CuO chains become superconducting via proximity effect. This leads to stronger Josephson coupling in c-axis direction and thus to the smallest superconductive anisotropy among all HTS families [67].

Beside these intrinsic obstacles for the transport of supercurrent in single-crystalline HTS materials there are additional hurdles since HTS materials are in general not a homogeneous continuum but rather a network of linked grains (see Fig. 5). The mechanism of crystal growth is such that all material that can not be fitted into the lattice structure of the growing grains is pushed forward into the growth front with the consequence that in the end all remnants of secondary phases and impurities are concentrated at the boundaries in-between the grains. Such barriers impede the current transport and have to be avoided by careful control of the growth process, in particular of the composition of the offered material.

Another obstacle for supercurrents in HTS is misalignment of the grains: Exponential degradation of the supercurrent transport is observed as a function of the misalignment angle (see Fig. 6). One of the reasons for this behavior is the d-symmetry of the superconducting order parameter (see Fig. 7) which meanwhile seems to be established for most of the HTS compounds [68]. The $J_c$ reduction as a function of the misalignment angle $\alpha$ turns out to be much larger than what is expected from d-wave symmetry only [69,70,71]. This extra $J_c$ degradation as well as the change of the current-voltage characteristics of the transport behavior[9] [75] are believed to arise from structural defects such as dislocations [76] and deviations from stoichiometry. In particular, the loss of oxygen at the grain surfaces [77] leads to a decrease of doping with respect to the grain bulk value and thus to a local degradation of the superconducting properties according to the temperature-doping phase diagram of HTS (see Fig. 8) [78]. Recently it has been demonstrated that for YBCO this effect can be partially eliminated by means of Ca-doping [79] which widens the range of acceptable grain misalignment in technical HTS material. In conclusion, manufacturing of technically applicable HTS materials requires well-defined preparation techniques to obtain chemically clean grains with small misorientation of their crystal axes.

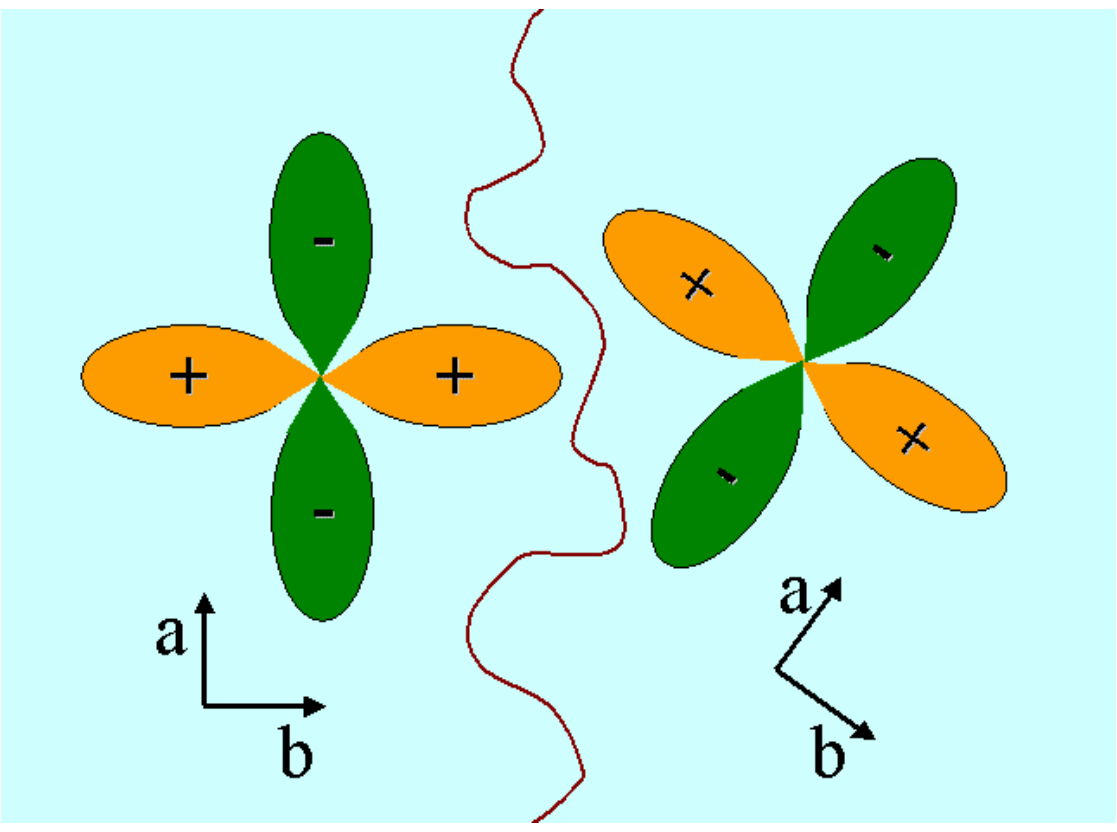

**Fig. 7** Schematics of a boundary between HTS grains. Misorientation of the superconducting d-wave order parameter leads to partial cancellation of the supercurrents modified by the faceting of the grain boundaries.

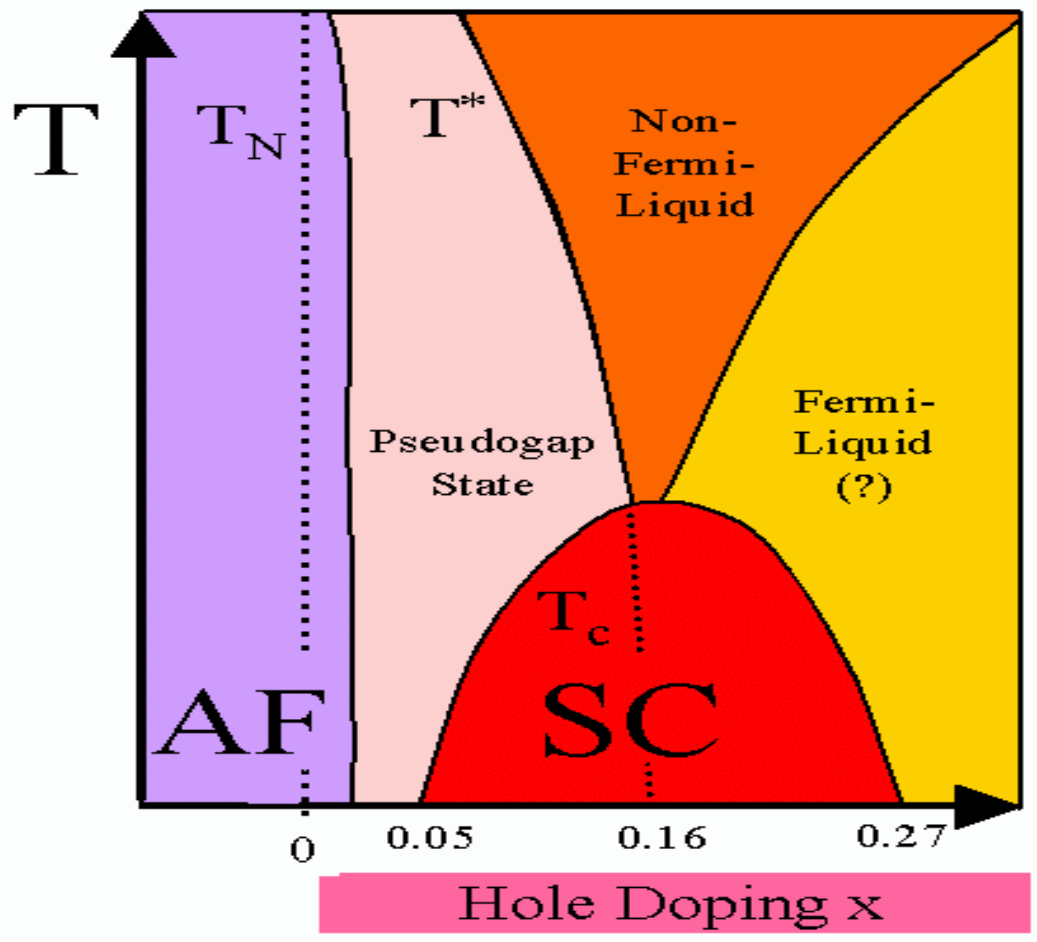

**Fig. 8** Schematic HTS temperature-doping phase diagram dominated by an interplay of antiferromagnetism ("AF") and superconductivity ("SC") [2,80,81].

[9] In addition to the quantitative effect on increasing misalignment angles of a reduction of the supercurrents, the current-voltage characteristics shows in addition a qualitative change from a flux-flow behavior for smaller misalignment angles [71,73] to overdamped Josephson junction behavior for larger misalignment angles [74].

## 4. Technically Applicable HTS Materials

The development of technically applicable HTS materials has progressed on several routes. Epitaxial HTS *thin films* achieve excellent superconducting properties ($T_c$ > 90 K; critical current density $J_c$ (77 K, 0 T) > $10^6$ A/cm$^2$; microwave surface resistance $R_s$(77 K, 10 GHz) < 500 μΩ, $R_s$(T,f) $\propto f^2$ ) that are well-suited for superconductive electronics. They are already in use in commercial and military microwave filter systems. HTS *Josephson junctions* based either on Josephson tunneling through ultra-thin artificial barriers or on the Josephson junction behavior of HTS grain boundaries have become available that can be used for the construction of highly sensitive magnetic field sensors ("Superconducting QUantum Interference Devices", "SQUIDs"). They are also tested for active electronic devices that can broaden the range of HTS thin-film applications. HTS single crystals are not suitable for applications due to their small size and their low $J_c$ values as a consequence of a low density of pinning centers. However, the "quick and dirty" version *melt-textured* HTS *bulk material* shows superb magnetic pinning properties and are already applied as high-field permanent magnets e.g. in magnetic bearings and motors. In spite of the ceramic nature of the cuprate oxides, flexible HTS *wire* or *tape conductor* material is obtained either by embedding HTS as thin filaments in a silver matrix or by HTS coating of metal carrier tapes.

Among these technical HTS materials, the Bi-HTS conductor material represents the only exception to the rule that strong biaxial texture is necessary to achieve technically relevant currents[10]. One reason is the exceptional softness of the Bi-22(n-1)n HTS as a consequence of the weak Van-der-Waals bonding between the two neighboring $BiO_x$ layers in the charge reservoir layers which allows alignment via mechanical processing steps like rolling or pressing. Another reason is the good mechanical and electrical contact with the Ag matrix due to similar melting temperatures which allows high current flow under the relaxed condition that a little detour via Ag is possible in case the direct current transfer between the HTS grains is blocked. This applies as well to "Ag-impregnated" Bi-2212 bulk which is commercially available in sizes of several 10 cm with homogeneous $J_c$ (77 K,0 T) of several kA/cm$^2$ [82,83].

Much higher critical current densities have been achieved in polycrystalline HTS materials, but their use implies a higher risk for applications. In all these cases, substantial $J_c$ reduction compared to the $J_c$(77 K,0 T) ~ 1 MA/cm$^2$ encountered in well-textured material with a sufficiently high amount of pinning centers always indicates a high degree of structural and electrical inhomogeneity with potentially devastating consequences: Forced current flow through defective materials regions may lead to a local quenching of superconductivity and the creation of "hot spots" [84]. The deposited quench energy leads then again to a structural and chemical modification of the HTS material in the neighborhood of these "hot spots". This enlarges the quench zone since such modifications are liable to spoil the conditions for good supercurrent transfer between neighboring HTS grains as discussed above. This spreading of the quench zone may finally end up in the destruction of the superconductor. This quench mechanism is different from the situation in classical superconductors where the quenching is caused by insufficient heat transfer due to the freezing of the phonon mechanism at LHe temperature. At the much higher operating temperature of HTS this is not such a critical issue since the heat distribution by phonons is here still very effective which brings about a comfortably high specific heat [85].

---

[10] Even though biaxial texture could be helpful in improving $J_c$ by up to one order of magnitude.

## 5. Thin Films

YBCO [86,87] and Tl-HTS (Tl-2212) [88] thin films are nowadays produced in quantities of several thousand wafers per year for commercial and military microwave filters. RE-123 films (RE = Nd, Sm, Gd, Dy,....) are investigated as YBCO alternatives with slightly higher $T_c$ values [89,90,91]. However, the required higher processing temperature increases the risk of cracks in the films [89]. Bi-HTS thin films are too soft for applications: High quality Bi-HTS films can be wiped away by cotton wool! Hg-HTS films are still restricted to small substrate sizes since fabrication in sealed quartz tubes is required in order to obtain good and stable films [92].

As for *YBCO*, *thermal coevaporation* has established as commercial *deposition method* [93, 94]. Substrates with sizes up to ∅ = 9" and 20 x 20 $cm^2$ can be coated on one side at a time at typical rates of 20 - 30 nm/min. Double-sided coating can be done in two successive deposition steps. Meanwhile, *pulsed laser deposition (PLD)* comes close to these possibilities [95]. *Magnetron sputtering* at deposition rates of up to 10 nm/min [91,96] is widely used in research laboratories. An extension to larger wafer sizes is still in preparation. *Molecular beam epitaxy (MBE)* has now also successfully demonstrated large area (DyBCO) deposition [97], but the much higher complexity of the deposition machines and lower deposition rates are intrinsic disadvantages. *Metalorganic chemical vapor deposition (MOCVD)* has the potential of deposition rates up to 1 µm/min [98], but the deposition of high quality YBCO films has only been reported for much lower rates. A special problem is the stability of the Ba precursor [99]. Recent quality improvement of YBCO thin films produced by *sol-gel* methods [100] is promising with respect to low cost production since the films can be prepared by fast non-vacuum deposition of precursor films with subsequent annealing to form the epitaxial films.

The higher $T_c$ > 100 K of *Tl-2212* allows a higher operation temperature than YBCO. Precursor deposition followed by an annealing step in $Tl_2O_3$ vapor is the regular fabrication method due to the high volatility of Tl [88]. Recently, improvement of the microwave properties has been achieved, however, at the expense of similar tendencies towards crack formation for increasing film thickness as in YBCO [101].

S*ubstrates* have to provide a suitably lattice-matched crystal matrix to align the HTS grains in uniform orientation [94]. Most suitable for applications is the c-axis orientation of the HTS films where the $CuO_2$ planes are parallel to the substrate which preserves the isotropy of the electrical properties in the film plane. Fortunately, this film orientation is promoted by the faster growth of HTS films along the $CuO_2$ plane directions compared to the c-axis direction. Matching of the thermal expansion coefficients of the substrates and the HTS films is an additional requirement since the HTS deposition takes place at 650 – 900 °C. Different contraction of HTS film and substrate on cooling to room or even cryogenic temperature leads to mechanical stress that can be tolerated by the film only up to a certain maximum thickness without crack formation [102]. This limits the thickness of YBCO films on the technically most interesting substrates silicon and sapphire to ~ 50 nm [103] and ~ 250-300 nm [104], respectively. Furthermore, these substrates have to be buffered e.g. by thin layers of $CeO_2$ and Y-stabilized $ZrO_2$ (YSZ) to prevent poisoning of superconductivity by diffusion of Si or Al into the HTS films. Buffer layers are also beneficial with respect to improved lattice match.

*Wet etching* [105] can be used for *patterning* of device structures with sizes down to the µm range. However, undercutting of the photoresist stencils causes problems in providing smooth sidewalls [106]. *Ar ion beam etching* is therefore the preferred method for the fabrication of µm size structures [106,107]. In combination with e-beam lithography and additional preparation techniques even sub-µm size structures can be realized [108,109].

While HTS single layer circuits have already reached a commercial stage of reproducibility, *multilayer circuits* can still be produced only at low yield. Due to the high volatility of Tl, multilayer technology with its need of high temperature deposition steps is only available for YBCO [110,111,112]. Recent improvement with respect to electrical insulation and high current capability in the different YBCO-layers as well as to the integrability of Josephson junctions gives rise to hope for an improved realization of HTS circuits with higher complexity [113,114,115,116].

## 6. Josephson Junctions

In HTS junctions, the d-wave symmetry of the superconducting wave function [117] and the vicinity of the metal-insulator transition leads to an intrinsically much more sensitive electrical transport behavior compared to classical superconductor junctions [118]: Bound Andreev states and the suppression of the order parameter at interfaces (depending on their orientation with respect to the crystal axes) introduce additional physical problems concerning spatial $J_c$ homogeneity. YBCO is the prevailing materials choice for HTS Josephson junctions due to its multilayer compatibility.

*Ramp-type junctions* [119] are still the most promising junction type for HTS digital circuits. The reproducibility of the contact parameters has reached a level which allows the production of digital circuits with a complexity of 10 – 100 junctions [121;115,116]. For previous barrier concepts based on thin insulating layers of (Ga-doped) PrBCO [109,119] or Co-doped YBCO [120,121] the $J_c$ homogeneity was limited by the growth control of this barrier layer on the ramp [119]. In "*interface-engineered*" junctions [115,116,123,124,125], instead of depositing a barrier layer a "natural barrier" (probably "pseudo-cubic" YBCO [126,127]) is formed by Ar ion etch treatment of the ramp base electrode [123]. Improved reproducibility and stability in subsequent high temperature treatments for multi-layer preparation have been reported [115,116]. Unfortunately, the barrier properties have turned out to show substantial dependence on the preparation conditions [124,125]. Another limiting factor is the geometrical uncertainty with respect to the ramp angle. Recent success in the fabrication of YBCO-Au-Nb ramp-type junctions [128] enables now the use of the d-wave symmetry related π phase shift as a new functional principle for the construction of novel electronic devices.

*c-axis microbridge (CAM) junctions* are c-axis interconnects between two superconducting layers separated by an insulator [129]. This is an attractive geometry, with low parasitic inductance suitable for multilayer circuits. Such junctions can be made by the planarization of a mesa or growth into a window in the insulator. The junction behavior is based on an Y rich interface prepared by the ion milling of the YBCO surface due to different milling rates of the cations. This places the mesa junctions in the class of interface engineered junctions with similar limitations. Geometrical uncertainties arise here from the preparation of submicron-size contacts.

*Step-edge junctions* [130,131,132] are still in use for 1- or 2-junction circuits such as SQUIDs [133]. Earlier hopes to control the growth behavior of the HTS films in the region of the substrate step in a sufficient way to achieve better reproducibility [130; 131] have not been fulfilled [134]. Stability of the junctions is an issue.

*SNS step-edge junctions* [135] with Ag-Au alloy as normal metal are also in use for SQUID magnetometers [137]. The point-contact-like nature of the transport from the normal metal to the YBCO electrodes and vice versa [135] indicates a stochastic flow of the supercurrent which gives little hope for a systematic improvement of the reproducibility of these contacts.

*Bicrystal junctions* [108,138,139,140,141,142] are widely used for demonstrator circuits [143,144,145]. They can easily be produced by single-film deposition followed by a simple photolithographic step at a reasonable degree of reproducibility. In multilayer technology even stacked junctions have been realized [113,143]. As for commercial circuits, the cost of the bicrystal substrates and the design restrictions imposed by the positioning of the contacts along the grain boundary appear to be prohibitive. The microscopic limit of $J_c$ homogeneity is given by the meandering of the grain boundary of the deposited HTS films which is connected intrinsically to the HTS film growth [146], and by the implications of d-wave symmetry for such a grain boundary geometry [117,147]. The reproducibility of junction parameters seems to be limited by defects in the substrate bicrystal grain boundary [148] which leads to an irregular course of its meandering.

*Biepitaxial contacts* [149,150,151,152] have slipped out of technological interest since only 45° [001] tilt grain-boundary contacts with low $J_c$ and $I_cR_N$ values have been developed to a more or less reproducible stage. Unfortunately, this geometry is a pathological case due to the d-wave symmetry of the superconducting wave function of HTS [153,154].

Junctions based on weak links introduced by *focused electron beam irradiation* ("FEBI") [155,156,157,158] have reached a high level of microscopic contact uniformity and have been used for the realization of single-layer digital circuits [159,160]. The contacts can be placed at an arbitrary position on the chip and allow in combination with focused ion-beam etching an absolute and relative positioning of the contacts with an accuracy of ~ 50 nm and ~ 1 nm, respectively. However, the strong temperature dependence of the $I_c$ values requires precise temperature stabilization. Due to the slow writing of the electron beam (~ 1 µm/min), each contact needs a "manufacturing time" of ~ 1 min. As for a similar approach based on ion implantation [161], compatibility of the junction fabrication with multilayer processing is unclear since irradiation of a YBCO layer will also affect buried layers and subsequent high-temperature process steps will affect already fabricated junctions by annealing.

Microbridges over substrate regions damaged by a *focused (Ga-) ion beam* ("FIB") [162,163] show Josephson contact like behavior that is based on the irregular growth behavior of the YBCO films on these substrate regions. Junctions with the desired RSJ-like IV-characteristic are therefore a rare exception [163].

There is still hope for classical *SIS sandwich contacts* once the oxide deposition will be controlled on an atomic scale [164,165,166].

Among non-YBCO contacts, *intrinsic Josephson junctions* in Bi-HTS [167,168,169, 170,171] or Tl-HTS [171] mesa structures are of interest with regard to rf oscillators and other active rf devices. The problem is not so much the preparation of the junctions that has already been realized in thin film technique [172] but the suitable microwave coupling arrangement that overcomes the impedance mismatch to conventional microwave circuits [173].

## 7. Wire & Tape Conductors

Bi-HTS /Ag tapes were the first practical HTS conductors [175] and have already reached a commercial stage: Tapes with cross sectional areas of ~ 1 $mm^2$ are available that can transport critical currents $I_c$ (77 K,0 T) up to 130 A [176] over km lengths. In the "Powder-in-Tube" fabrication, Ag tubes are filled with Bi-HTS precursor powder and are subject to various thermomechanical processing steps where the tubes are flattened out to tapes that include the Bi-HTS material as thin filaments. The fragility of the Bi-22(n-1)n grains along neighboring Bi-oxide planes[11] is the physical starting-point for the successful mechanical alignment of the Bi-HTS grains by means of rolling or pressing which orients the grains preferably with the ab-planes parallel to the tape surface. Silver is the only matrix material that is chemically not reacting with HTS compounds and allows for sufficient oxygen diffusion. The common melting of Ag and Bi-HTS around ~ 850° C gives rise to a close mechanical and electrical contact that helps to bridge non-superconducting regions by means of low-resistivity shorts. However, the cost and the softness of Ag are critical issues for technical applications.

Among the two "practical" Bi-HTS systems Bi-2212 and Bi-2223, Bi-2212 has the advantage of a much better control of the chemical phase development [177,178]: Bi-2212 forms from a single phase precursor whereas Bi-2223 requires a multi-phase starting mixture. Today, a quite homogeneous superconductive connection of the HTS grains can be achieved in Bi-2212 filaments [179] with *engineering critical current densities* (calculated with respect to the total conductor cross section) $J_{eng}$ (4.2 K, 0 T) > 70 kA/$cm^2$ over several 100 m conductor length [180]. By clever arrangement of Bi-2212 tapes, $J_{eng}$ (4.2 K, 24 T) > 20 kA/$cm^2$ has been achieved in a practical wire conductor independent of the direction of the applied magnetic field. By new processing techniques Bi-2212 tapes have been fabricated with *superconducting critical current densities* $J_c$ (4.2 K, 10 T) > 500 kA/$cm^2$ (calculated with respect to the HTS part of the conductor cross section) already close to the limit of epitaxial thin films [181]. This demonstrates good perspectives for Bi-2212 conductor with respect to applications in high magnetic fields at low temperature [182].

Bi-2223/Ag tapes were for a long time in the center of interest of HTS conductor development since their $T_c$ > 100 K allows operation at $LN_2$ temperature [183,184,185]. However, since the phase evolution of Bi-2223 is much more complicated compared to Bi-2212, the pronounced percolative superconductive current flow in the Bi-2223 filaments could not yet be overcome [186]. Moreover, the upper field limit of ~ 1 T at 77 K [187] restricts the $LN_2$ operation of Bi-2223 tapes to low-field applications such as cables [188,189] or transformers. For operation at technically interesting magnetic field levels of several Tesla, Bi-2223 windings have to be cooled to $T_{op}$ < 30 K. However, in this temperature region Bi-2212 conductors may be competitive due to simper processing and thus lower fabrication cost [181].

Cost arguments have always been in favor of HTS coating of simple robust metal carrier tapes. However, the transfer of YBCO thin film techniques to such a conductor fabrication has taken a lot of efforts since biaxial texture without any interruption is required over the whole conductor length. Meanwhile, $J_c$ (77 K, 0 T) ~ 1 MA/cm$^2$ is achieved over a length of ~ 10 meters which allows first practical demonstrations. This status of conductor length development is comparable to the situation of Bi-HTS conductors 10 years ago.

[11] The two $BiO_x$ layers in the charge reservoir layers $SrO/(BiO_x)_2/SrO$ of the Bi-HTS are attached to each other only by weak van-der-Waals-like bonding.

Tl-1223 [190] and Hg-1212 [191] are under investigation with respect to HTS coatings with $T_c$ > 100 K. However, the difficulties of the required thin film techniques suggest that these HTS are at present no practical alternatives to YBCO. Earlier efforts with respect to a fabrication of Tl-1223 conductors by means of processing techniques developed for Bi-HTS conductors have also not been successful [192].

## 7. A. Bi-2212

Bi-2212 tape conductors with a HTS fill factor of up to 60 % can be produced by simple surface coating of Ag bands. By improved processing ("Pre-Annealing and Intermediate Rolling", "PAIR") tapes with $J_c$ (4.2 K, 10 T) > 500 kA/cm$^2$ have been fabricated [181]. Trade-off between $J_c$ and the thickness of the Bi-2212 layers [193] limits the expectations for $J_{eng}$ (4.2 K, 20 T) of optimized long tapes to 50 - 60 kA/cm$^2$ [194].

As practical demonstration, such tape conductor was assembled in a cable similar to a "Rutherford cable", the standard conductor for accelerator magnets: It consisted of 427 Bi-2212 filaments in a AgMgSb alloy outer sheath to allow for higher stress tolerance compared to pure Ag [195]. This wire exhibited excellent $I_c$ (4.2 K, self-field) = 285 A and $J_c$ (4.2 K, self-field) = 340 kA/cm$^2$ and a room temperature tensile strength of 180 MPa. For a full cable $I_c$ (4.2 K, self-field) = 3500 A was obtained in a short sample, $I_c$ (4.2 K, self-field) > 2700 A in 17-m-long sections cut from a continuous 80-m-long cable. Beside a possible application for next generation accelerator magnets, the cable is also investigated with respect to its use in Superconducting Magnetic Energy Storage devices ("SMES") [196].

By clever arrangement of differently oriented Bi-2212 tapes[12], a wire conductor with high $J_{eng}$ (4.2 K, 24 T) > 20 kA/cm$^2$ independent of the direction of the applied magnetic field has been constructed which is suitable for the fabrication of conventional solenoid magnets ("ROtation-Symmetric Arranged Tape-in tube wire", "ROSATwire") [180,197]. The Bi-HTS fill factor could be increased to ~ 40 % [198].

This development aims at insertion magnets for 30 T class magnets where the Bi-HTS coils have to generate an additional field of several Tesla in a background field of ~ 20 T [199]. The most ambitious goal is the use of such magnet systems for high-field NMR: The extreme requirements of temporal stability can probably only be fulfilled in a superconducting magnet system operated in persistent mode which relies on an extremely slow decay of the supercurrents [200]. However, the demonstrated decay times are still not yet sufficient to achieve this goal.

[12] The ROSATwire conductor is an assembly of an equal number of tapes rotated around the tape axis by 0, 120 and 240 degrees, respectively. In an external magnetic field, the field component perpendicular to the ab-planes and thus to the tape surface has the strongest effect on $J_c$. This results for the ROSATwire tape arrangement in a $J_c$ suppression which is nearly independent of the conductor orientation.

## 7. B. Bi-2223

Bi-2223 Powder-in-Tube tapes have seen for more than ten years a continuous performance increase. The critical current densities in the HTS filaments were raised up to $J_c$ (77 K, 0 T) ~ 70 kA/cm$^2$ in short samples [201]. Better understanding of the mechanical deformation process [202] and of the Bi-2223 phase development during thermal annealing [203] helped to establish a production of several 10 km of such tape conductors with $J_{eng}$(77 K, 0 T) ~ 15 kA/cm$^2$ at a total conductor cross section of ~ 1 mm$^2$ [183,184,185]. Locally, $J_c$ (77 K, 0 T) up to 180 kA/cm$^2$ has been observed [186], but this seems to be the upper limit for Bi-2223/Ag tapes [204]. The reason are non-superconducting inclusions and pores in the HTS filaments leading to a percolative supercurrent flow along the Bi-2223 grains which are well connected but make up at most only ~ 2/3 of the filament cross-section [205]. With an HTS fill factor of high-$J_c$ Bi-2223 conductors of at most 35 % an intrinsic limit $J_{eng}$ (77 K, 0 T) ~ 60 kA/cm$^2$ has to be expected which is still comparable to the limit of ~ 100 kA/cm$^2$ estimated for YBCO coated conductors (assuming a future development of $J_c$ (77 K, 0 T) ~ 1 MA/cm$^2$ in 5 μm thick YBCO coatings on 50 μm thick metal carrier tapes).

However, for applications in higher magnetic field a substantially lower operating temperature is required with the additional benefit of a $J_c$ increase of up to a factor of 7 compared to 77 K ($J_c$(T, 0 T) ≈ $J_c$(77 K, 0 T) [7 – T/13 K] [206]). Another economical disadvantage with respect to YBCO coated conductors is the mandatory use of silver as matrix material. Previous cost estimations of 10 $/kA m for full-scale production levels [201] (at present: 200 - 300 $/kA m) have turned out to be too optimistic [207] and are meanwhile adjusted to 25 -50 $/kA m in 2006. Moreover, while the good electrical contact of the HTS filaments with the Ag matrix is helpful in bridging disconnected HTS regions by means of low resistance shorts [208], frequent use of this current rerouting prevents persistent-mode operation of Bi-2223 coils.

Bi-2223 conductors have overcome many problems on the way to a technical HTS material. The problem of the softness of the Ag matrix has been solved by applying dispersed MgO in the Ag matrix [209] or by adding a thin layer of stainless steel reinforcement to both sides of the tapes which enables the tapes to withstand a tensile stress of 300 MPa and a tensile strain of 0.45% at 77 K [176]. Promising approaches have been developed to tackle the problem of ac losses which arise from electromagnetic coupling of the HTS filaments [210,211] and are of concern for all power applications since they determine the necessary cooling power [212,213,214,215,216]: Insulating $BaZrO_3$ [212] or $SrCO_3$ barriers [213] around the HTS filaments and higher resistive Ag alloys such as AgAu or AgPd as matrix material as well as twisting of the filaments (down to twist pitches $l_p$ ~ 1 cm [213]) reduces these coupling currents. A novel wire arrangement of horizontal and vertical stacks of Bi-2223 tapes reduces the $I_c$-anisotropy with respect to the orientation of an external magnetic field [217].

In conclusion, Bi-2223 conductors have arrived at a practical level of technical applicability, however, still at a quite high cost level. At present, the biggest psychological handicap for Bi-HTS conductors are the great expectations for a soon arrival of a cheaper and better YBCO coated conductor.

## 7. C. YBCO

YBCO-coated metal tapes are promising with respect to lower-cost HTS conductor operating even at $LN_2$ temperature in magnetic fields up to several Tesla. $J_c$ (77 K, 0 T) = 1 - 2 MA/cm$^2$ has been achieved in short samples by a number of YBCO coating methods choosing different routes to biaxial YBCO texture. One route is the deposition of a textured buffer layer on (untextured) metal with the assistance of an *ion beam* that *introduces orientation-selective growth*. With stainless steel as attractive choice for the metal, 10 m long high-$J_c$ tapes have been demonstrated [218,219]. Another route is the use of *cube-textured metal bands*. Buffer layers ($CeO_2$,Y-stabilized $ZrO_2$ (YSZ),$Gd_2Zr_2O_7$ (GZO), In-Sn oxide (ITO), …) have to be applied here as well in order to compensate the lattice mismatch and to prevent poisoning of superconductivity in the YBCO coating by in-diffusion of metal atoms. Ni and Ni alloys are used here as metal substrates that show the required cube-texture after appropriate metallurgical and heat treatment.

Most of the tested buffer materials are insulators with the consequence that a disruption of the superconducting current path can not be bridged by a low-resistance short via the metal tape as in the case of Bi-HTS/Ag tapes. Practical solutions of this problem are the deposition of a Au layer on top of the YBCO coating [218] or use of metallic oxide buffer layers [220].

$J_c$ (77 K, 0 T) ~ 1 MA/cm$^2$ is obtained in YBCO coatings with a thickness of up to ~ 3 µm [218,221]. For ~ 100 µm thick metal carrier tapes this results in a HTS fill factor of only ~ 3 %. The engineering current density $J_{eng}$ calculated with respect to the total conductor cross-section of $J_{eng}$(77 K, 0 T) ~ 30 kA/cm$^2$ is therefore still close to what is achieved in present commercial Bi-2223/Ag tapes. 50 µm thick metal carrier tapes are available. Ca doping and optimization of the grain architecture may help to transfer the high intra-grain $J_c$ (77 K, 0 T) ~ 5 MA/cm$^2$ [222] into macroscopic current densities even for the case of a grain alignment of only 10° [223].

By *Ion Beam Assisted Deposition* ("*IBAD*") [221,224,225] tapes with $J_c$ (77K, 0 T) = 2.2 MA/cm$^2$ ($I_c$ (77K, 0 T) = 78 A) measured over a length of 10 m [218] and $J_c$ (77K, 0 T) = 0.5 MA/cm$^2$ over a length of 46 m have been fabricated. The deposition of the required ~ 1 µm thick YSZ buffer under assistance of an additional ion beam selecting the required biaxial texture is painfully time-consuming. This problem can be overcome by MgO-IBAD-buffering where a buffer thickness of ~ 10 nm is already sufficient for biaxial texture [226]. Meanwhile, $J_c$ (77K, 0 T) = 0.9 MA/cm$^2$ ($I_c$ (77K, 0 T) = 14.4 A) has been achieved with this technique over a length of 0.8 m [230]. This renders again the YBCO laser deposition as the time-limiting fabrication step at present coating rates of up to 60-70 nm × m$^2$/h [218]. The cost of the laser deposition is an economical issue: IBAD-YBCO tapes are still more expensive than Bi-2223/Ag tapes [218].

*Inclined Substrate Deposition* ("*ISD*") achieves biaxial alignment of the YSZ buffer without assistance of an additional ion-beam using a high rate laser deposition and appropriate inclination of the metal substrate with respect to the laser plume [227]. This makes the YSZ buffer deposition by a factor 10 faster than IBAD. However, the degree of biaxial alignment is not yet sufficient: $J_c$ (77 K, 0 T) ~ 1 MA/cm$^2$ has been achieved up to now only in short samples [228].

In *Rolling-Assisted-Biaxially-Textured-Substrates* ("*RABiTS*"), cube-texture of a Ni (alloy) substrate tape is generated by conventional rolling with heavy deformation (> 95 %) to a roll textured tape, followed an annealing step which results in a recrystallization into the desired biaxially textured cubic phase [229,231]. Sophisticated buffering techniques have been tested in order to avoid or remove the oxide layer from the tape surface since it destroys the unique biaxial orientation of the substrate and may blast the YBCO coating due to the volume expansion accompanying the oxidation. With a $CeO_2$/YSZ/$CeO_2$ buffer system $J_c$ (77 K, 0 T) > 1 MA/$cm^2$ has been achieved in small samples [229], but only $J_c$ (77 K, 0 T) ~ 0.6 MA/$cm^2$ over 0.8 m tape length [230].

"*Surface-Oxidation Epitaxy*" ("*SOE*") is a similar approach where cube-textured pure Ni tape is used [232] which is oxidized under controlled condition to form cube-textured NiO. The actual texture "seen" by the YBCO coating is here not the Ni texture but the texture of the NiO which is suitably lattice matched to YBCO. $J_c$ (77 K, 0 T) ~ 0.1 MA/$cm^2$ demonstrated in short samples with additional MgO buffering is not yet satisfactory.

*Biaxially textured Ag-substrates* are investigated for YBCO coating as well [233]. However, the use of expensive and soft Ag carrier tapes in combination with the low HTS fill factor of the tape coating approach is not very promising. Moreover, the achieved degree of textured grains is still well below the level of nearly 100 % which is already obtained for Ni and Ni alloy tapes.

Substantial cost-reduction can be expected if less costly deposition methods than the technically successful but expensive vacuum physical vapor deposition approaches can be employed. Recent progress in *solution-based* YBCO *coating* (dip coating, spray pyrolysis, sol-gel) with first $J_c$ (77 K, 0 T) ~ 1 MA/$cm^2$ samples is most promising [234,235]. Recently, MOD-YBCO coated tapes with $I_c$ (77 K, 0 T) = 140A over 7.5 m tape length have been reported. *Liquid Phase Epitaxy* is another promising option [236].

## 8. Bulk Material

Melt-textured YBCO pellets [237] with good superconducting properties ($J_c$ (77 K, 1 T) > 10 kA/$cm^2$, $J_c$ (50 K, 10 T) > 100 kA/$cm^2$) can be grown reproducibly in sizes up to a diameter d = 6 cm even in complex shapes [238] comparable to what has been achieved in NdFeB permanent magnets. The present production cost of > 1000 EUR / kg (including quality control) is still a factor of ~ 10 higher than for NdFeB [239] but it can be expected to come closer this cost level once a production of several tons per year can be established here as well. Sizes up to d = 10 cm have been realized [240] but joining of smaller size pellets with good superconducting properties of the connections [241] seems to be a more practical solution[13]: For larger pellets full oxidation becomes increasingly difficult even though it is not based on oxygen bulk diffusion [243] but on oxygen transport along microcracks [244,245].

[13] The product of $J_c$ and sample size r = d/2 is probably the most appropriate quantitative criterion for the superconducting quality of bulk samples since it is proportional to the magnetic field trapping capability [242]. Even if the $J_c$ of joints is at present still substantially below the $J_c$ of high-quality bulk samples a higher $J_c \times r$ product can be obtained by joining such samples.

The strong pinning allows the "freezing" of high magnetic fields. In single domain cylinders with a diameter of 30 mm an induction of 1.3 T can thus be fixed at 77 K [238], 16 T have been achieved at 24 K in-between two d = 2.2 cm samples [246], very recently even 17 T at 29 K in-between two d = 2.65 cm / h = 1.5 cm samples [247]. This exceeds the potential of conventional permanent magnets based on ferromagnetic spin polarization by an order of magnitude and can be used for high-field "cryomagnets". The present field limit of YBCO cryomagnets does not stem from the pinning, but from the involved mechanical forces which exceed at high fields the fracture toughness of the YBCO ceramic [248] (20 - 30 MPa in YBCO without additions [249]) and leads without appropriate mechanical stabilization to cracking of the pellets. Ag addition [246,250] were shown to be helpful in providing higher intrinsic stability. Passivation by resin impregnation [248,251] protects the samples against humidity and thus improves the long-term stability.

The combination of melt-textured YBCO with permanent magnets leads to levitation properties that are attractive for bearing applications. In contrast to classic magnetic bearings, superconducting magnetic bearings are self-stabilized due to flux pinning. The strong levitation forces (e.g. > 80 N for zero field cooled d = 3 cm YBCO pellets at 77 K at a distance of 0.5 mm from a 0.4 T / d = 2.5 cm SmCo magnet [238]) with force densities of > 10 $N/cm^2$ and the stiffness are already limited by the magnetic field strength of the permanent magnet [252,253,254]. YBCO pellets are even under consideration as substitute for LHe-cooled NbTi-coils in the Japanese Maglev project [255].

In contrast to YBCO, high-quality rare-earth 123 samples can not be fabricated in air but require oxygen-reduced processing atmosphere. For melt-textured Nd-123 [256,257] and Sm-123 [258] improved superconducting properties have been reported but only for sample sizes up to d ~ 3 cm. Nevertheless, the freezing of 2.0 T maximum induction in d ~ 5 cm sized Gd-123 samples at 77 K and trapping of 3.3 T in between two such samples at 77 K [259] is beyond what has been achieved for YBCO. Ternary mixtures of Rare-Earth-123 [260] have demonstrated even better pinning of stronger magnetic field at 77 K [261].

Bi-2212 rods and tubes are commercially available in sizes up to d = 30 cm and 50 cm length with electrical properties sufficient for many applications such as current leads or fault current limiters [82].

## 9. Conclusion

In sixteen years since their discovery, High-Temperature Superconductors proceeded quite far towards the goal of a materials basis from which engineers can choose well-specified products to build technical systems. However, the HTS spectrum of such industrial feedstock is still limited. Nevertheless, from the experience gained in the development of these materials it has become clear that certain promised benefits can be expected from a technical use of HTS within the next couple of years.

## Acknowledgement

I would like to thank W. Goldacker, T. Habisreuther, M. Lakner, G. Linker, T. Scherer, C. W. Schneider, P. Seidel, M. Siegel, and T. Wolf for critically reading the manuscript and for many helpful comments.